\documentclass[3p,times,procedia]{elsarticle}
\usepackage{nupha_ecrc}
\usepackage{wrapfig}
\usepackage{subcaption}
\usepackage{lineno} 

\volume{00}

\firstpage{1}

\journalname{Nuclear Physics A}

\runauth{}

\jid{nupha}

\jnltitlelogo{Nuclear Physics A}

\usepackage{amssymb}

\usepackage[figuresright]{rotating}

\begin{document}

\begin{frontmatter}



\dochead{}

\title{Longitudinal Asymmetry and its Measurable Effects in Pb-Pb Collisions at 2.76 TeV}


\author{Rashmi Raniwala (for the ALICE Collaboration)}
\address{Physics Department, University of Rajasthan, Jaipur, India; email:rashmi.raniwala@cern.ch}

\begin{abstract}
Collisions of identical nuclei at finite impact parameter have an unequal
number of participating nucleons from each nucleus due to
fluctuations. The event-by-event fluctuations have been estimated by 
measuring the difference of energy in the zero-degree calorimeters on
either side of interaction vertex. The fluctuations
affect the global variables such as the rapidity distributions, and
the effect has been correlated with a measure of these fluctuations.
\end{abstract}


\end{frontmatter}


\section{Introduction}
\label{}

In a collision of two identical nuclei at any impact
parameter, density fluctuations lead to an unequal number of participants
from each colliding nucleus. The participating nucleons make up the 
participant zone, which has a non-zero momentum in the nucleon-nucleon
centre-of-mass (CM) frame, which is also the laboratory
frame for most collider experiments. These collisions have an innate
longitudinal asymmetry which is expected to manifest itself in some
measurable observables. A limiting case of such an asymmetry is seen
in p-A collisions where there is an excess of particles on the A-side
as compared to the p-side, and the pseudorapidity distribution may be a little wider
on the A-side. This asymmetry will affect all measurements that are
affected by fluctuations, by the number of participants and by the
symmetry about the nucleon-nucleon CM frame. One such observable is the rapidity
distribution of charged particles. In the present talk, we will
estimate the initial state asymmetry and find its effect on the
rapidity distribution by comparing them in collisions of different asymmetries.

\begin{figure}[h]
\vspace{-10pt}
\centering
\includegraphics[width=0.7\textwidth]{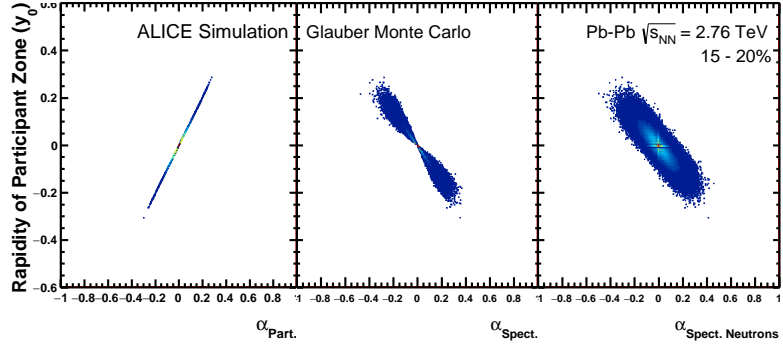}
\caption{The rapidity shift of the participant zone as a function of
  the asymmetry in
  (i) the number of participants (ii) the number of spectators and (iii)
  the number of spectator neutrons for 15-20\% centrality, as obtained in
  a Glauber Monte Carlo.}
\label{fig:y0asym}
\vspace{-15pt}
\end{figure}

The initial state asymmety is defined using the generic definition
$\alpha_{\rm {part}} = (A-B)/(A+B) $, where $A$ and $B$ are the number of
participating nucleons from each nucleus. This asymmetry causes a
shift in the rapidity of the participant zone $y_{\rm 0} = \frac{1}{2} {\rm ln}
\frac{A}{B} = \frac{1}{2} {\rm ln}\frac{1+\alpha_{\rm {part}}}{1-\alpha_{\rm
    {part}}}
$. At an impact parameter of 6.5 fm,  Glauber model ~\cite{Cons}
calculation gives the mean number of participants in a Pb-Pb collision
as 120.  Assuming fluctuations to cause a collision to
have 126 and 114 participants from the two nuclei
yields $y_{\rm 0} \approx 0.05$.  The
shift in rapidity of the CM is related to the asymmetry in the number
of participants, in the number of spectators, and in the number of
neutron spectators, albeit with a poorer resolution, as shown in the
three panels of Fig. ~\ref{fig:y0asym}. 

Data was recorded for Pb-Pb collisions at $\sqrt{s_{\rm {NN}}}$ =2.76 TeV in 2010 in
the ALICE experiment. The
centrality of the collision was estimated using the multiplicity in the
V0 detectors ( $ -3.7 < \eta < -1.7 $ and $ 2.8 < \eta < 5.1 $) and also using the track multiplicity in
the Time Projection Chamber (TPC) and the Inner Tracking System (ITS)
in the region $-0.8 < \eta < 0.8 $.
The effect of asymmetry was studied on the charged particle pseudorapidity distributions measured using the hits in the V0
detector and also using the charged particle tracks constructed using
information from the TPC and the ITS ~\cite{PerfAlice}. The analysis
was performed on a total of about 3 M events in 0-40\% central events,
equally populating 5\% centrality bins. The data
corresponded to a
minimum bias trigger with the vertex cuts $|V_z| \leq
5$ cm and $|V_x,V_y| \leq
0.3$ cm.  The spectator neutrons are measured in the
neutron Zero-Degree Calorimeters (ZDCs) on either side of the interaction
vertex ($\eta >$ 8, at about 114 meters away from the nominal
interaction vertex).
\section{Analysis: Measuring Asymmetry and its Effect on (Pseudo)rapidity Distributions}
For the most central collisions the spectator size is small, and
fragments into individual nucleons.  The few neutrons measured in the ZDC are a
good estimate of the neutron spectators, and hence of the asymmety of
the collisions. With increasing impact parameter,  the larger spectator
fragments into smaller nuclei and individual nucleons. The charged
spectators are bent away from the 
direction of neutron ZDCs by the magnetic field. The neutrons reaching
the ZDCs are less than the actual number of neutron spectators.  The
effect of neutron loss on the asymmetry is investigated in simulations. 
\begin{figure}[h]
\begin{subfigure}{0.5\textwidth}
\vspace{-10pt}
\centering
\includegraphics [trim=1.0cm 0.0cm 0.0cm 0.0cm,clip,
width=0.75\linewidth]{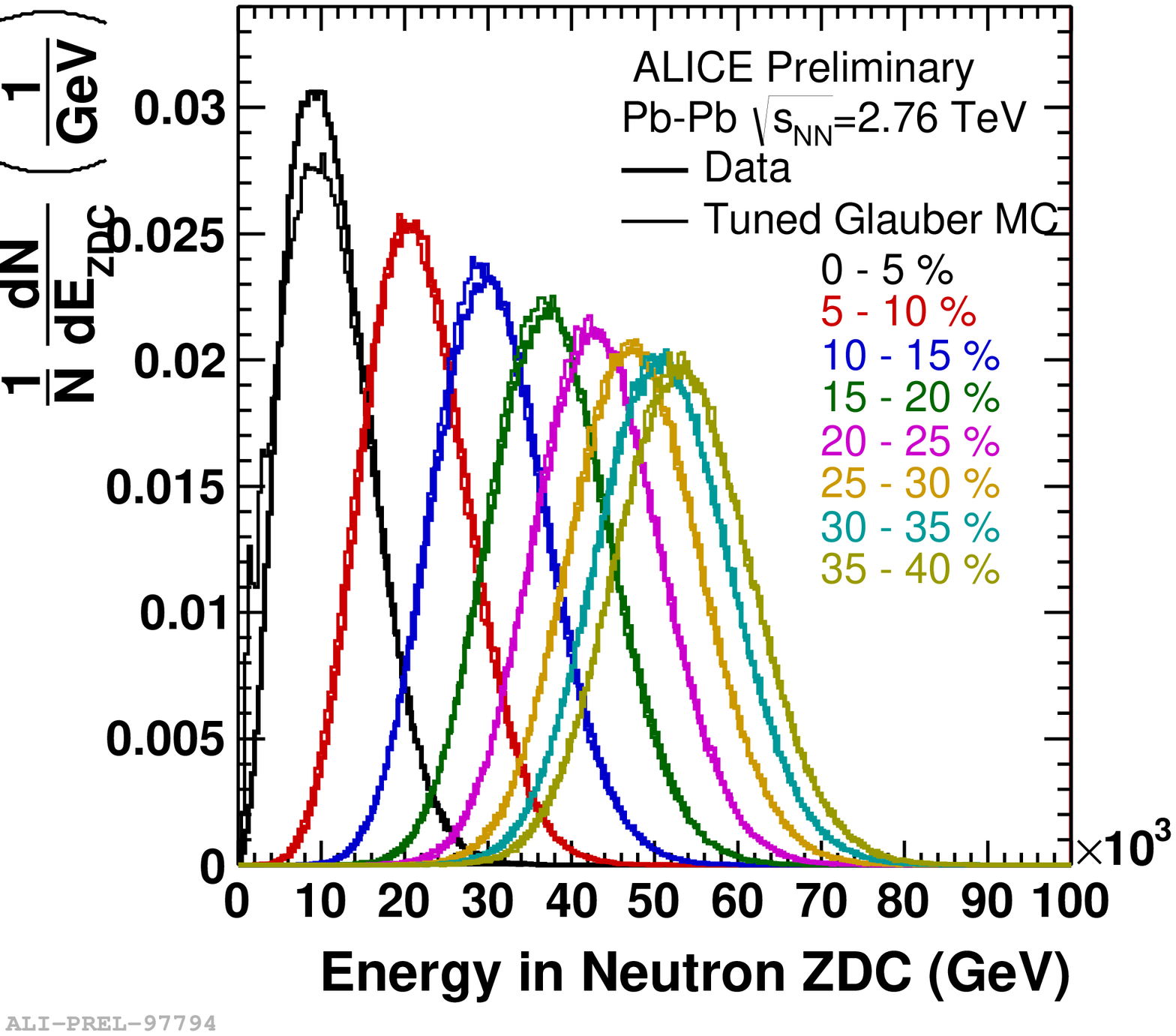}
\vspace{-5pt}
\caption{}
\vspace{-8pt}
\end{subfigure}
\begin{subfigure}{0.5\textwidth}
\vspace{-10pt}
\centering
\includegraphics [trim=0.0cm 0.0cm 0.0cm 0.0cm,clip,
width=0.75\linewidth]{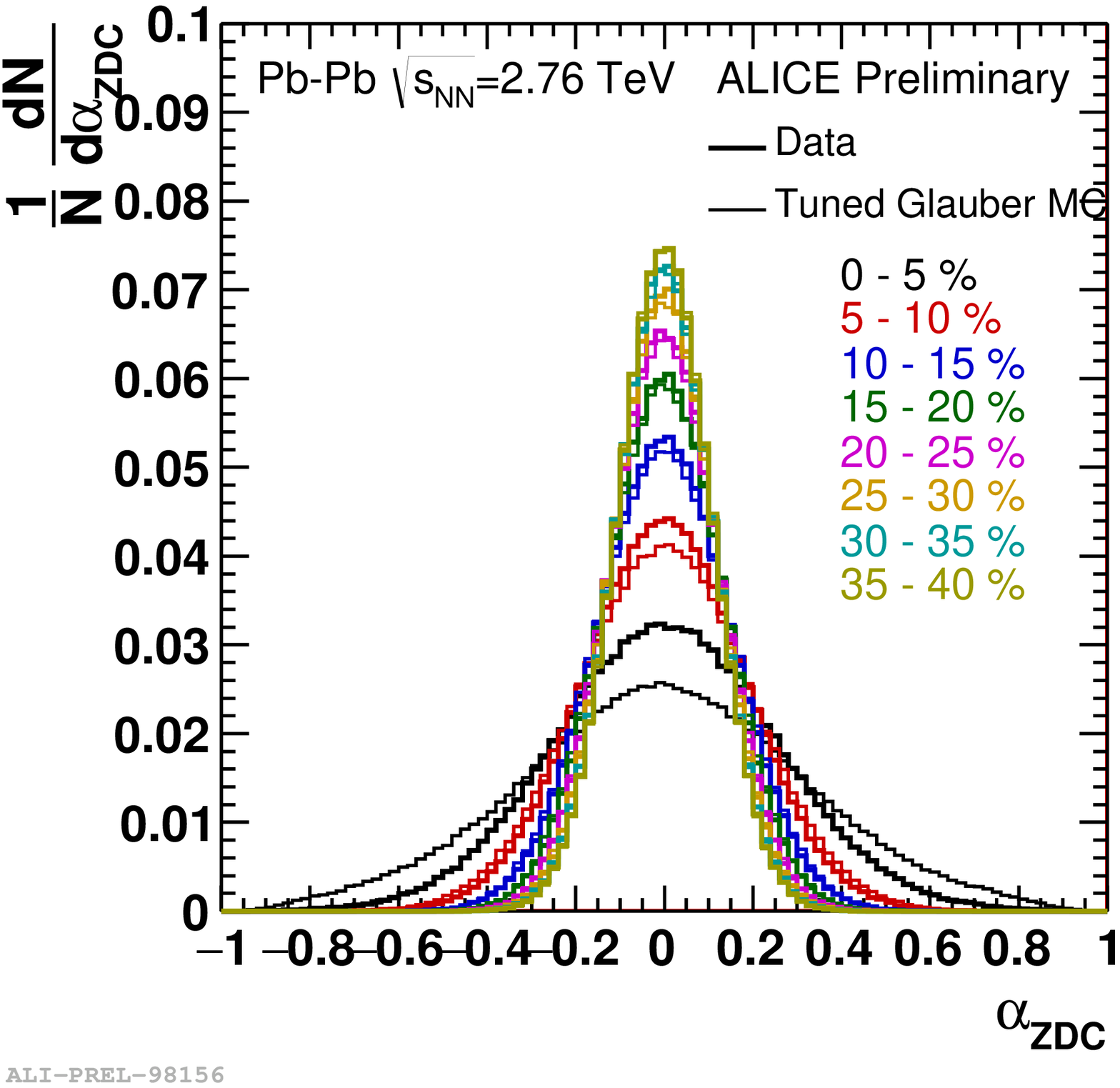}
\vspace{-5pt}
\caption{}
\vspace{-8pt}
\end{subfigure}
\caption{(a) The energy distribution in the neutron ZDC in each 5\%
  centrality interval for TGMC and for measured data (b) The
distribution of the asymmetry parameter $\alpha_{\rm {ZDC}}$ in simulation and
in data for different centralities.}
\label{fig:ZDC_Energy}
\vspace{-8pt}
\end{figure}

Knowing the number of neutron spectators and the response of the ZDC ~\cite{CentAlice},
the energy distribution in the ZDCs was obtained
in a Tuned Glauber Monte-Carlo model (TGMC) by tuning the loss of neutrons using a common
functional form to reproduce the
energy spectrum in the ZDCs for different centralities. 
Fig.~\ref{fig:ZDC_Energy} shows the distribution of energy in the ZDC,
and the distribution of asymmetry, as obtained using the tuned Glauber
model and in data. The asymmetry in each event is obtained as
$\alpha_{\rm {ZDC}}
= \frac {ZNC-ZNA}{ZNC+ZNA} $ where $ZNC$ and $ZNA$ denote the energy
measured in that event in the two neutron ZDCs on either side of the
interaction vertex. The agreement for all cases, other than the most
central, is quite good, validating the use of ZDC measurement as an
estimate of the asymmetry.  

Experiments over the last many decades have shown that particle
production is symmetric about the CM in N-N collisions, as also
expected from the symmetry of the collision. In nucleus-nucleus collisions,
assuming that the particle production is symmetric about the shifted
rapidity of the participant zone, and assuming that the particle rapidity distribution can be described
by a Gaussian, the ratio of ${\rm d}N/{\rm d}y$ distributions for asymmetric and symmetric
events can be expressed as 
\begin{equation}
\frac {({\rm d} N/{\rm d}y)_{\rm asym}}{({\rm d}N/{\rm d}y)_{\rm sym}}  =
\frac {N exp (-\frac {(y-y_{\rm 0})^2}{2\sigma_{\rm y}^2})} {N exp
(-\frac {y^2}{2 \sigma _{\rm y}^2})} 
\end{equation}
For symmetric collisions, $y_{\rm 0} = 0.$ Considering that $y_{\rm 0}$ is
small, the ratio yields $ \approx  1 + \frac {y y_{\rm 0}}{\sigma_{\rm
    y}^2}
$. Since the experiment measures the pseudorapidity distributions,
it is worth investigating whether the ratio remains linear with
pseudorapidity.  Toy-model simulations suggest that the ratio can be
approximated as 1 + $c_{\rm 1}\eta$, and the
coefficient $c_{\rm 1}$ is still related to the shift in the CM rapidity by $y_0
\approx c_{\rm 1} \sigma_{\rm \eta}^2$, to within 5\%.

\begin{wrapfigure}{r}{0.5\textwidth}
\vspace{-13pt}
\centering
\includegraphics[width=0.4\textwidth]{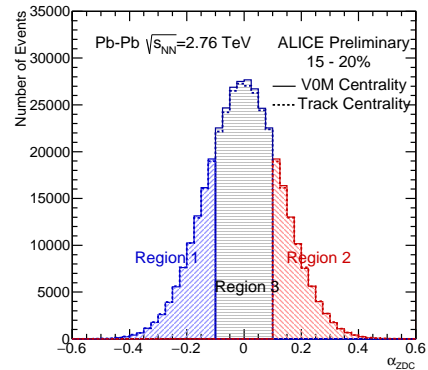}
\caption{Distribution of the asymmetry $\alpha_{\rm {ZDC}}$ in the 15-20\%
  centrality interval.}
\label{fig:ZDCAsyReg}
\vspace{-10pt}
\end{wrapfigure} 
The asymmetry distribution of each centrality class may be used to
classify the events as  symmetric or asymmetric events.
 Fig. ~\ref{fig:ZDCAsyReg} shows the asymmetry distribution as obtained using
measurements in the ZDCs, for the centrality
interval 15-20\%. Events with $ -0.1 < \alpha_{\rm {ZDC}} < 0.1 $
(region 3) are
considered as symmetric events and events with $|\alpha_{\rm {ZDC}}| > 0.1$ were
considered as the two Regions (1 and 2) of asymmetric events.

\begin{figure}[h]
\begin{subfigure}{0.5\textwidth}
\vspace{-10pt}
\centering
\includegraphics[width=0.9\textwidth]{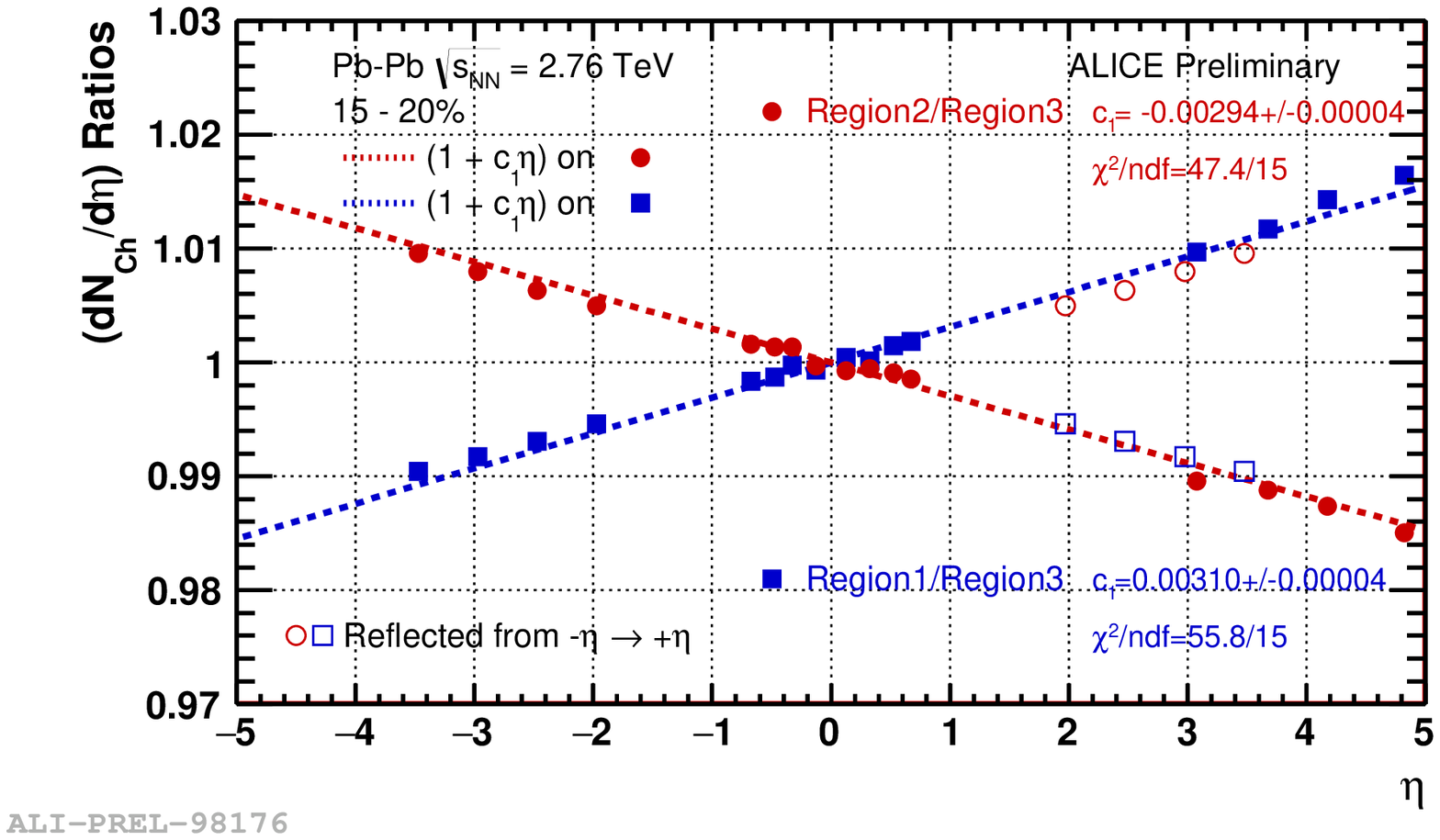}
\vspace{-5pt}
\caption{}
\vspace{-8pt}
\end{subfigure}
\begin{subfigure}{0.5\textwidth}
\vspace{-10pt}
\centering
\includegraphics[width=0.9\textwidth]{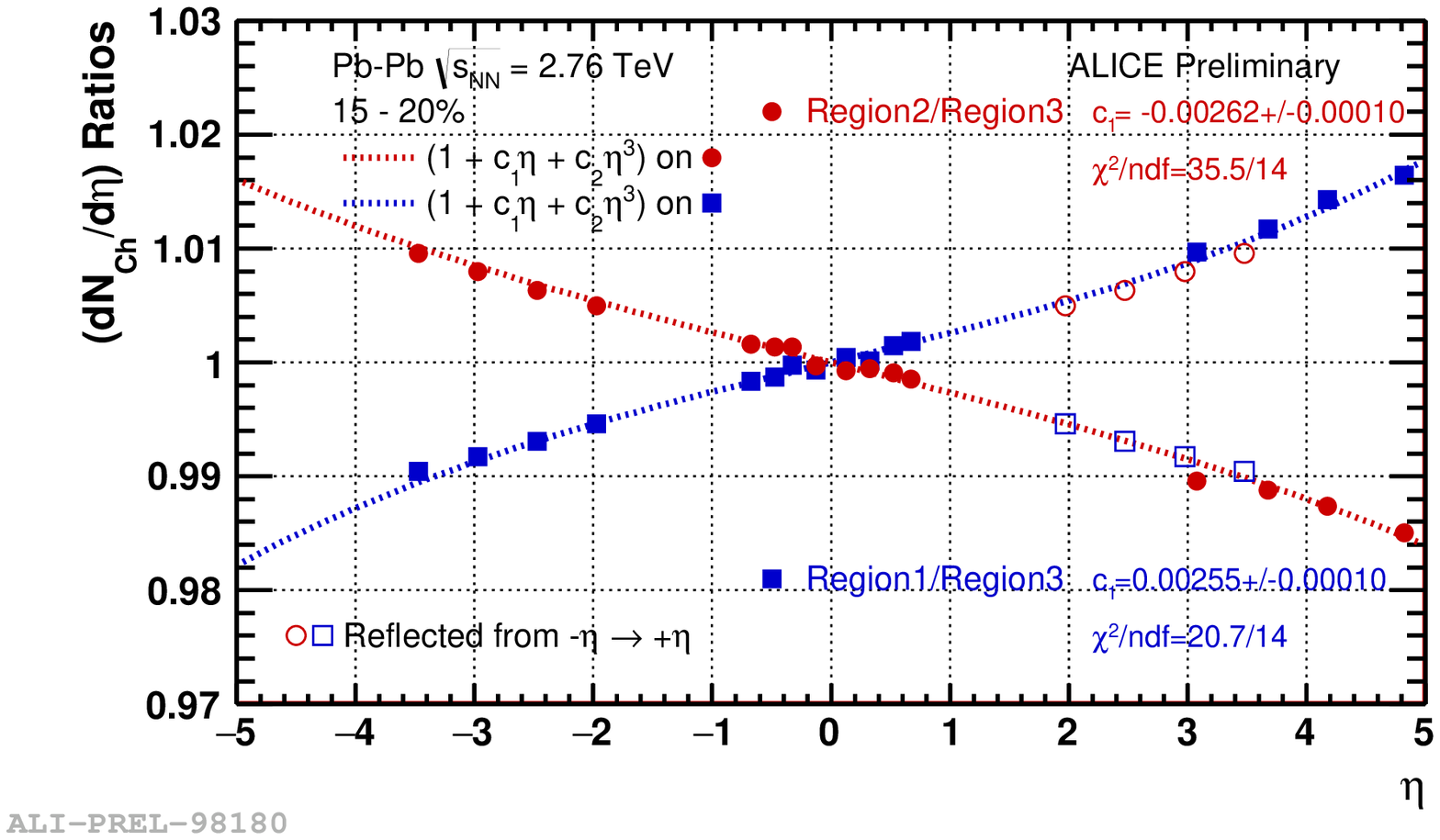}
\vspace{-5pt}
\caption{}
\vspace{-8pt}
\end{subfigure}
\caption{(a) The figure shows the fits to the ratio of dN/d$\eta$. The
  data points for $\eta < $ 0.9 are from TPC and ITS reconstructed
  tracks and for $\eta > $
  1.0 are from hits in the V0A and V0C. The open symbols are reflected points and
  not used in the fit. (a) Data fitted with 1 +
  $c_{\rm 1}\eta$ (b) Data fitted with 1 +
  $c_{\rm 1}\eta$ + $c_{\rm 2}\eta^3$.  In both cases, the value of the
  coefficient $c_{\rm 1}$ and the $\chi^2/$NDF are shown on the plot. The
  cubic function provides a better fit to the data. }
\label{fig:dndeta1}
\end{figure}

The
ratio of the measured pseudorapidity distributions for Region-1 to
Region-3 and
Region-2 to Region-3 can be used to obtain the coefficient
$c_{\rm 1}$. Taking the ratio of the pseudorapidity distributions cancels
out most systematic uncertainties. The ratio of ${\rm d}N/{\rm d}\eta$ is obtained in
the central rapidity region, and in the region of acceptance of the V0 detectors.
The linear fits (1 + $c_{\rm 1}\eta$) to the ratios are shown in
Fig. ~\ref{fig:dndeta1}(a). The details are described in the caption.
The values of the coefficients and the $\chi^2/$NDF
are shown in Fig.~\ref{fig:dndeta1}.
Since the goodness of fit
estimates are not too good, a cubic term ($c_{\rm 2}\eta^3$) was added and the resulting
fits are shown in Fig.~\ref{fig:dndeta1} (b).
\section{Results}
\begin{figure}[]
\begin{minipage}{0.5\textwidth}
\vspace{-10pt}
\centering
\includegraphics[width=0.9\textwidth]{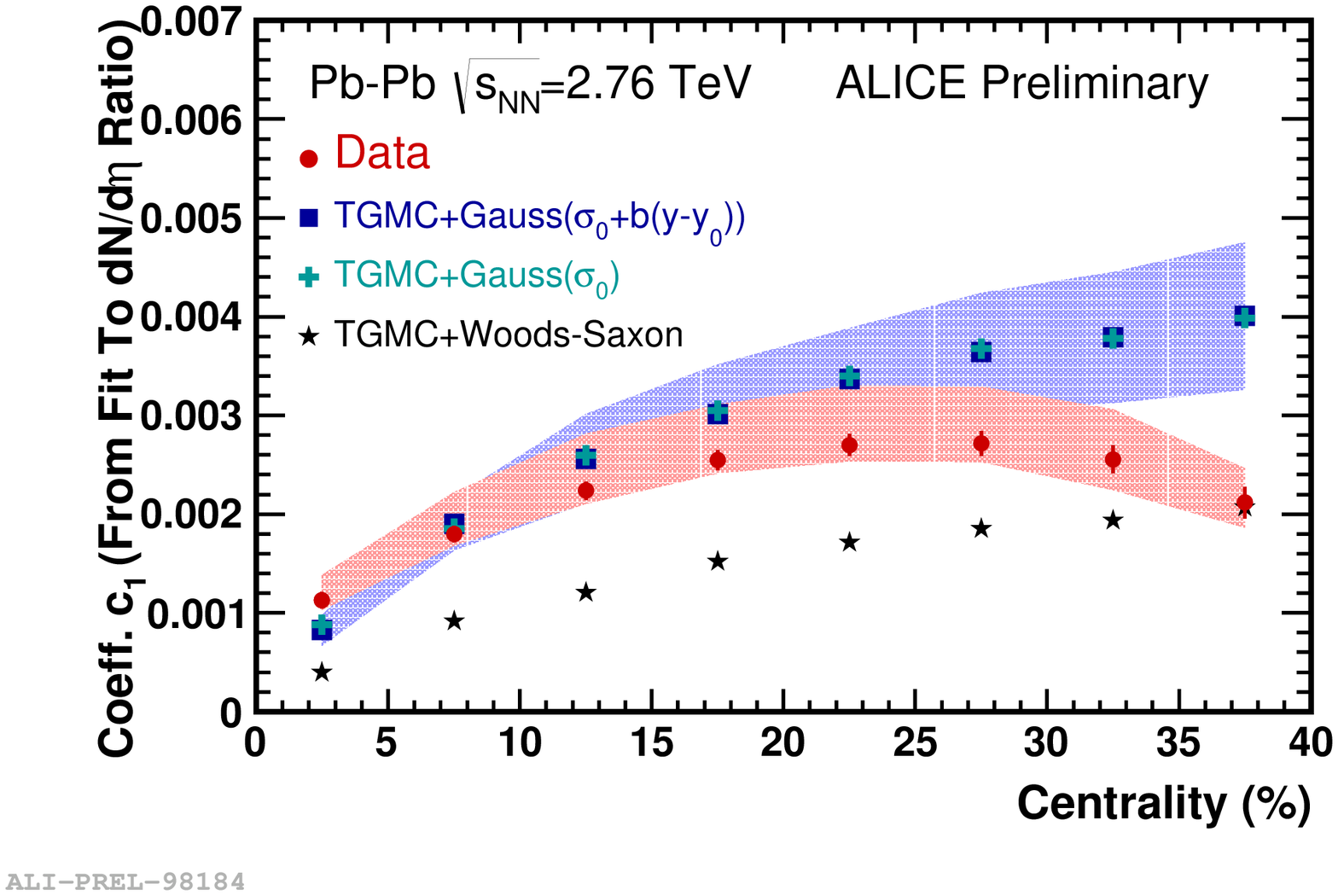}
\caption{Coefficient $c_{\rm 1}$ for different centralities, as described in
text.}
\label{fig:c1f}
\end{minipage}
\begin{minipage}{0.5\textwidth}
\vspace{-10pt}
\centering
\includegraphics[width=0.9\textwidth]{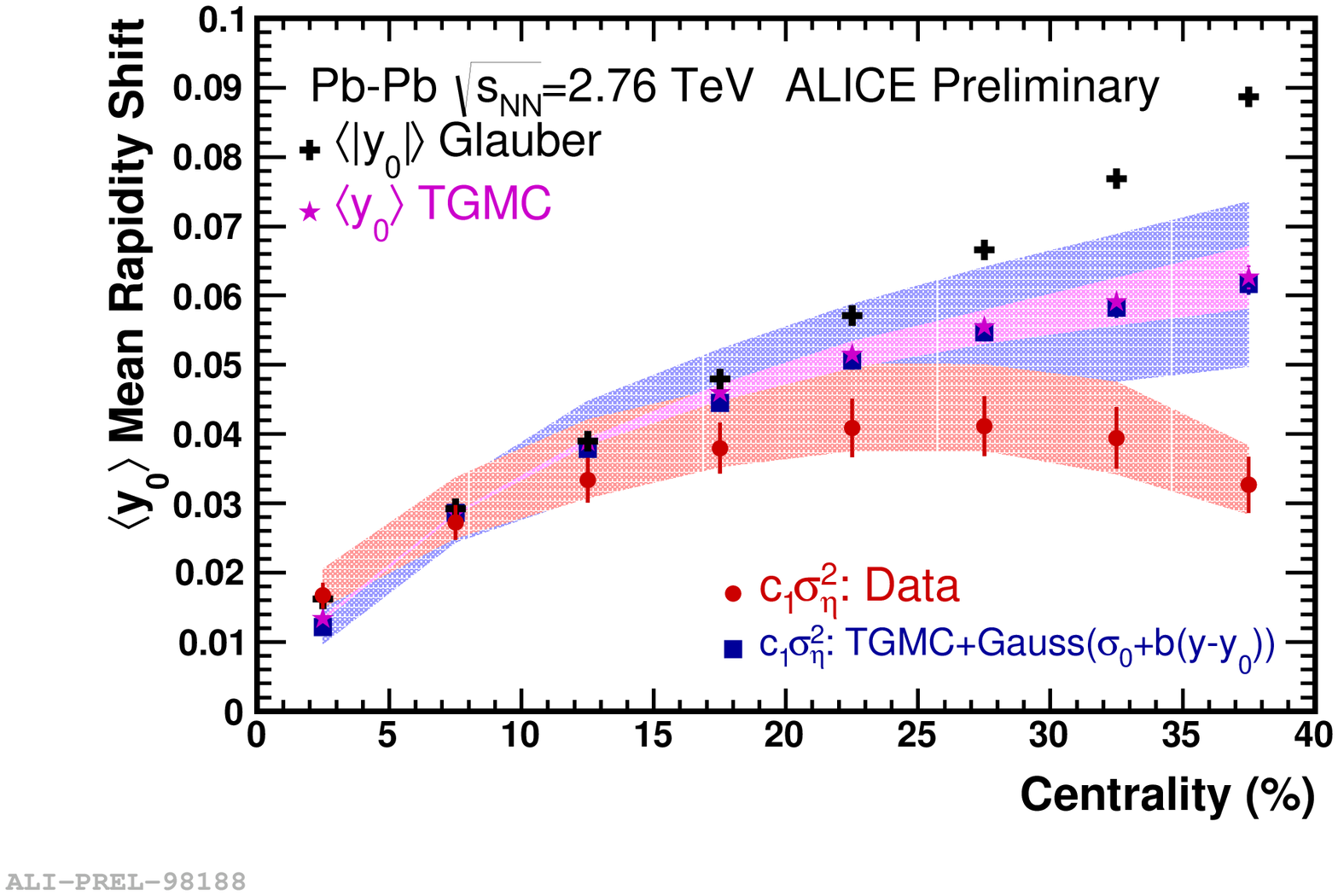}
\caption{Rapidity shift $y_{\rm 0}$ for different centralities.}
\label{fig:y0est}
\end{minipage}
\vspace{-10pt}
\end{figure}

The ratios of ${\rm d}N/{\rm d}\eta$ distributions for each centrality were fitted
with both the linear function and the cubic function for both
Region-1/Region-3 and Region-2/Region-3.
The cubic function fitted
better in each case. The coefficient $c_{\rm 1}$ is plotted as a function of
centrality of the collisions in Fig. ~\ref{fig:c1f}.
The systematic errors
shown in the figure are due to changes in centrality critieria, vertex
cuts, difference in Region-1 and Region-2, and difference in values of
$c_1$ in a linear and in a cubic fit.

Using the TGMC, a response matrix was
obtained between shift in the rapidity of the participant zone
($y_{\rm 0}$) and the
event-asymmetry obtained from simulated values of energy deposited in 
neutron ZDCs, similar to the response matrix from neutron spectators
shown in the last panel of Fig.~\ref{fig:y0asym}. The distributions of
$y_{\rm 0}$ were obtained for
each asymmetry region by unfolding the corresponding matrix. The
process was repeated for each centrality.  
The resulting $y_{\rm 0}$-distributions were used  along with different assumptions of initial rapidity
distributions (Gaussian, Gaussian-like with rapidity dependent sigma, Woods-Saxon), and realistic {\it p}$_{\mathrm T}$ distributions
~\cite{CentRatios} in a toy-model simulation to determine the
${\rm d}N/{\rm d}\eta$ distributions for different asymmetry regions. The free parameters in the initial rapidity
distributions were chosen to reproduce the experimental pseudorapidity
distributions. The ratio of ${\rm d}N/{\rm d}\eta$ for asymmetric to symmetric
region (Region-1(2)/Region-3) was fitted with a cubic function. The
resulting values of $c_{\rm 1}$ are also shown in
Fig.~\ref{fig:c1f}.

Fig.~\ref{fig:y0est} shows the values of the rapidity-shift $y_{\rm 0}$
obtained in four different ways, including the known values from the  
 Glauber model calculation. Using the TGMC, the mean value of rapidity
 shift was obtained for the asymmetric events as described
 above. The empirical relation between $c_{\rm 1}$ and the rapidity shift
($y_{\rm 0} = c_{\rm 1} \sigma_{\rm {\eta}}^2$) is used for the data and for the
simulated events. For more central events, there is good agreement between the values of
$y_{\rm 0}$ obtained in different ways.
\section{Summary}
The present analysis demonstrates that fluctuations in the initial
state cause a longitudinal asymmetry which shifts the rapidity of the
participant zone by $y_{\rm 0}$. The asymmetry has been estimated by
measuring the energy in the 
ZDCs on either side of the interaction vertex, providing a model
dependent method for determining the shift in rapdity $y_{\rm 0}$. This
asymmetry was shown to manifest itself in the final state
pseudorapidity distributions of charged particles and the magnitude of the effect is related
to the initial state asymmetry. The event-asymmetry is a useful
variable to classify events to investigate the effect of longitudinal
fluctuations and to study the systematic dependence of various
observables on this asymmetry.












\end{document}